\documentclass[ reprint, amsmath, amssymb, aps, prl]{revtex4-2}

\usepackage{graphicx}% Include figure files
\usepackage{dcolumn}% Align table columns on decimal point
\usepackage{bm}% bold math
\usepackage{siunitx}
\usepackage{ulem}
\begin{document}

\title{Experimental Observation of Spontaneous Emission of Space-Time Wavepacket in~a~Multimode Optical Fiber} 

\author{Karolina Stefańska$^{1,2}$}
\author{Pierre Béjot$^{1}$}
\author{Karol Tarnowski$^{2}$}
\author{Bertrand Kibler$^{1}$}
 \email{bertrand.kibler@u-bourgogne.fr}
\affiliation{
 $^{1}$Laboratoire Interdisciplinaire Carnot de Bourgogne (ICB), UMR6303 CNRS-Université Bourgogne Franche-Comté, 21000 Dijon, France\\
 $^{2}$Department of Optics and Photonics, Wrocław University of Science and Technology, Wybrzeże Wyspiańskiego 27, 50-370 Wrocław, Poland
}
\date{\today}

\begin{abstract}
We provide a complete analysis, from theory to experiment, of the spontaneous emergence of a~discretized conical wave of X-type (i.e., a localized 2D+1 space-time wavepacket) when an intense ultrashort pulse nonlinearly propagates in a multimode fiber. In particular, we reveal that this spatiotemporal phenomenon corresponds to broadband intermodal dispersive wave emission from an unsteady localized wave structure formed during nonlinear propagation. Theoretical phase-matching predictions are experimentally and numerically confirmed in a commercially-available step-index multimode fiber. Our results provide a general understanding of phase-matched radiations emitted by nonlinear waves in multidimensional dispersive optical system.

\end{abstract}

\maketitle

\section{Introduction}

Multimode optical fibers (MMFs) offer an enabling platform for investigating spatiotemporal phenomena in multidimensional systems and complex collective dynamics. This simply lies in the possible fine control and analysis of transversal modes that can propagate and interact, in contrast to limiting cases of single-mode waveguides and bulk media with their infinite and continuous set of modes. Over the past few years very exciting observations have been reported in terms of novel optical effects, such as geometric parametric instability, beam self-cleaning, light thermalization processes, multimode solitons, and spatiotemporal laser mode-locking, to name a~few \cite{picozzi2015nonlinear,wright2015controllable,krupa2016observation,krupa2017spatial,renninger2013optical,wright2017spatiotemporal,pourbeyram2022direct,podivilov2022thermalization}.
When combining the potential of multimode waveguides with recent technological advances in laser sources and pulse shaping, one of the most promising development relates to the generation and application of propagation-invariant spatiotemporal (ST) wavepackets \cite{kibler2021discretized,bejot2021spatiotemporal,shiri2020hybrid,bejot2022quadrics,shiri2022propagation}. Such an ultimate control of light will pave the way to the delivery of pulsed light with custom spatial profiles through multimode optical waveguides for sensing, imaging, and spectroscopy, and to novel light-matter interactions. Until now, invariant optical ST wavepackets have been mainly generated and characterized in free space and bulk media \cite{kondakci2019optical,yessenov2021space,yessenov2022space} and often limited to 1D+1 configurations (e.g., ST light sheets).

In this contribution, we demonstrate the spontaneous generation of a discretized conical wave from nonlinear pulse propagation in a multimode fiber. This spatiotemporal phenomenon that leads to the formation of a 2D+1 linear wavepacket results from a coherent emission of intermodal dispersive waves driven by a characteristic and non-strict phase-matching relation. Even if the resulting ST wavepacket is not fully non-dispersive and non-diffractive, our work appears as a welcome alternative to the exact and complex shaping of invariant multidimensional ST wavepackets for future investigations.

\section{From theory to experiment}

Conical waves are simple examples of ST wavepackets that have been shown to spontaneously emerge during the nonlinear propagation of ultrashort and ultraintense laser pulses in bulk media \cite{faccio2007conical,hernandez2008localized}. The conical emission generally observed during the filamentation process turns out to be a particular example of conical waves emergence, i.e. a manifestation of X-wave generation in the normal dispersion regime, thus taking the form of a~hyperbola in the $(k_{\perp};\omega)$ plane, where $k_{\perp}$ is the transversal component of the wave vector and $\omega$ the temporal frequency. Recently, it was numerically suggested that MMFs can also support spontaneous conical emission during the propagation of ultrashort pulses with peak powers around the critical self-focusing threshold of silica glass \cite{bejot2019multimodal,kibler2021discretized,tarnowski2021numerical}.
In contrast to bulk medium, MMFs support only a discrete number of guided modes at a~given frequency $\omega$. Consequently, the modal distribution of optical fibers provides a discretization of conical emission (e.g., discretized X waves). This arises from the linear superposition of fiber modes with an engineered spatiotemporal spectrum given by the following phase-matching relation:
\begin{equation}
    \Big|\beta_{(m,\omega)}-\beta_{(1,\omega_0)}-\left[\beta_{1}+\delta\beta_1\right]\left(\omega-\omega_0\right)\Big|\leqslant\frac{2\pi}{d_z},
\label{eq:conicalemission}
\end{equation}
\noindent where $\beta_{(m,\omega)}$ refers to the full frequency-dependent propagation constant of guided mode $m$ ($m=1$ corresponding to the fundamental mode) obtained with a scalar mode solver, $\omega_0$ is the pump central frequency, and $1/\beta_{1}=v_\mathrm{g0}$ is the group velocity of the fundamental mode at $\omega_0$.
In other words, the generated spacetime wavepacket propagating at the group velocity $v_\mathrm{gX} = 1/(\beta_1+\delta\beta_1)$ is composed by the family of modes respecting Eq.\,\ref{eq:conicalemission} and, in the case of normal dispersion, forms a discretized X-pattern in the $(m,\omega)$ plane \cite{kibler2021discretized}.Parameter $\delta\beta_1$ indicates the difference in the usual retarded time frame given by the group velocity $v_\mathrm{g_{0}}=1/\beta_{1}$ of the fundamental mode at $\omega_0$, which is explicitly: $\delta v_g = \delta\beta_1 / \left[\beta_1\left(\beta_1 + \delta\beta_1\right)\right]$.
The particular case of spontaneous conical emission refers to the above non-strict phase-matching condition (Eq.\,\ref{eq:conicalemission}) since it takes place over a finite distance $d_z$ seeded by a~broadband (ultrashort) wave structure (with group velocity $v_\mathrm{gX}$), thus implying a certain tolerance about the bandwidth of generated frequencies \cite{kibler2021discretized,kolesik2005interpretation}. 
The resulting conical wave is not fully non-dispersive and non-diffractive. Only an exact linear multiplexing of monochromatic modes that satisfies the above relation (Eq.\,\ref{eq:conicalemission}) with strict equality could overcome such limitations.

Our numerical approach of nonlinear pulse propagation in MMFs is based on the multimode unidirectional pulse propagation equation (MM-UPPE) \cite{bejot2019multimodal}, which describes the
evolution of the complex electric field $\bar{A}$ in the scalar approximation limited here to axially symmetric linearly polarized modes (LP$_{0,m}$ modes).
This equation written in the modal basis takes form:
\begin{equation}
\begin{aligned}
&\partial_z\bar{A}(m,\omega)=i\left(\beta(m,\omega)-\frac{\omega-\omega_0}{v_{g_{0}}}\right)\bar{A}
+\frac{i\,n_{\textrm{eff}_0}n_2\omega^2}{c^2 \beta(m,\omega)}&\\
&\left\{\left(1-f_R\right)\overline{|A|^2A}
+f_R\overline{\left[\int h_R(\tau)|A(t-\tau)|^2d\tau\right]A}\right\},&
\end{aligned}
\label{EqPropComplex}
\end{equation}
where $n_{\textrm{eff}_0}$ is the effective refractive index of the fundamental mode at the central frequency $\omega_0$ of the laser pulse, $n_2$ is the nonlinear refractive index of the medium (here for silica glass, we used $n_2 = 2.6\times10^{-20} \mathrm{m}^2/\mathrm{W}$). The function $h_R$ is the Raman response of the fiber medium with Raman fraction $f_R = 0.18$ for fused silica glass. The propagation equation is expressed in a~retarded frame moving at velocity $v_\mathrm{g_{0}}$.

\begin{figure}
    \centering
    \includegraphics[width=0.95\linewidth]{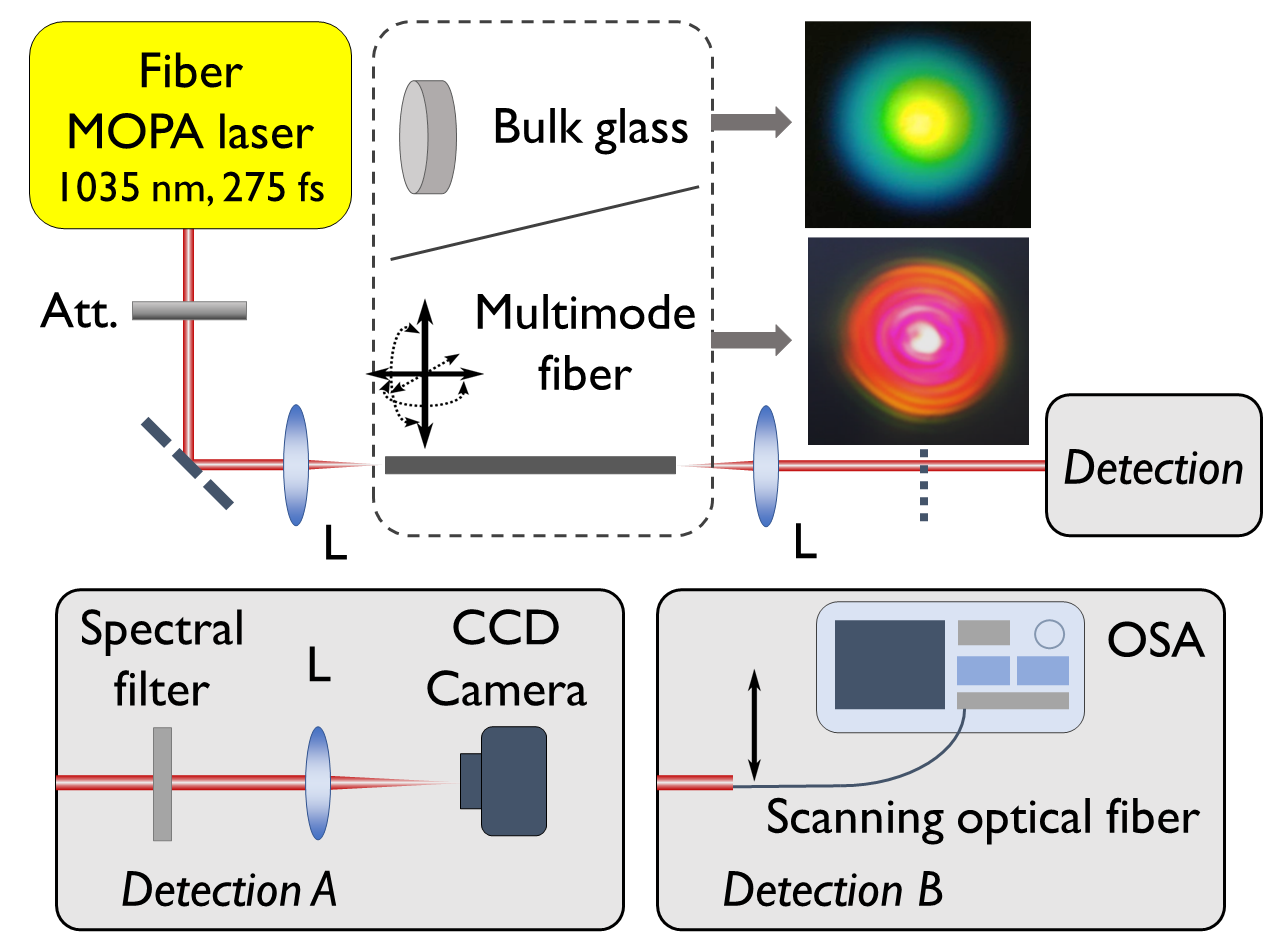}
    \caption{\label{fig:1} Experimental setup for capturing near-field images and recording angularly-resolved far-field spectra of spontaneous conical emission from a multimode fiber and bulk glass. Att.: variable attenuator; L: lens; OSA: optical spectrum analyzer. The insets show photos of the output beam in the far field in both cases.}
\end{figure}

To reveal the existence of discretized conical emission in a MMF, we implemented the following experimental setup depicted in Figure~\ref{fig:1}. We used a high-power femtosecond laser (Monaco 1035, Coherent) delivering \SI{275}{ \femto\second} pulses at \SI{1035}{ \nano\meter} with a chosen repetition rate of \SI{10}{ \kilo\hertz}. The pulse energy can be as high as \SI{60}{ \micro\joule}. The laser output power was adjusted to  \si{\micro\joule}-level energy by means of neutral density filters and a half-wave plate combined with a polarizer, in order to prevent damage issues \cite{ferraro2022multiphoton}. The high-quality beam was carefully coupled into the fiber with a $f=\SI{12.5}{\centi\meter}$ lens (Thorlabs, LB1904-C), through an optimization of the beam diameter at the focal point (nearly $\SI{45}{ \micro\meter}$) to only excite the fundamental guided mode. We used a commercially-available step-index multimode fiber (Thorlabs, FG105LCA) with a pure silica core (core diameter $\Phi= \SI{105}{\micro\meter}$) and numerical aperture $\text{NA}=0.22$. The length of the fiber was \SI{5.8}{\centi\meter}. We placed the fiber segment on a multi-axis platform that provides 5 degrees of freedom to ensure optimal coupling into the fundamental mode. We then analyzed light at the fiber output by means of a beam profiler with a CMOS sensor (CinCam CMOS-1202 IR, Cinogy) and an optical spectrum analyzer (Yokogawa AQ6374), in particular for capturing near-field images (detection system A) and recording angularly-resolved far-field spectra (detection system B). 

In addition, we carried out the experimental observation of conical emission in a bulk glass sample in a similar way. We used a sapphire plate (5-mm thick), in order to unveil similar and distinct features between both systems. In both cases, the propagation takes place in the normal dispersion regime. From simple pictures of the output beams in the far field shown in Fig.~\ref{fig:1}, one can easily observe the typical frequency-angular intensity distribution of the field, for the visible spectral tail. In both cases, the highest frequencies are observed over increasing angles (i.e., larger $k_{\perp}$). However, in the case of the MMF, we notice discrete rings, instead of a continuous concentric rainbow, which are formed by the superposition of the fiber modes with increasing order that contain the higher frequencies.

\section{Results}

\begin{figure}
    \centering
    \includegraphics[width=\linewidth]{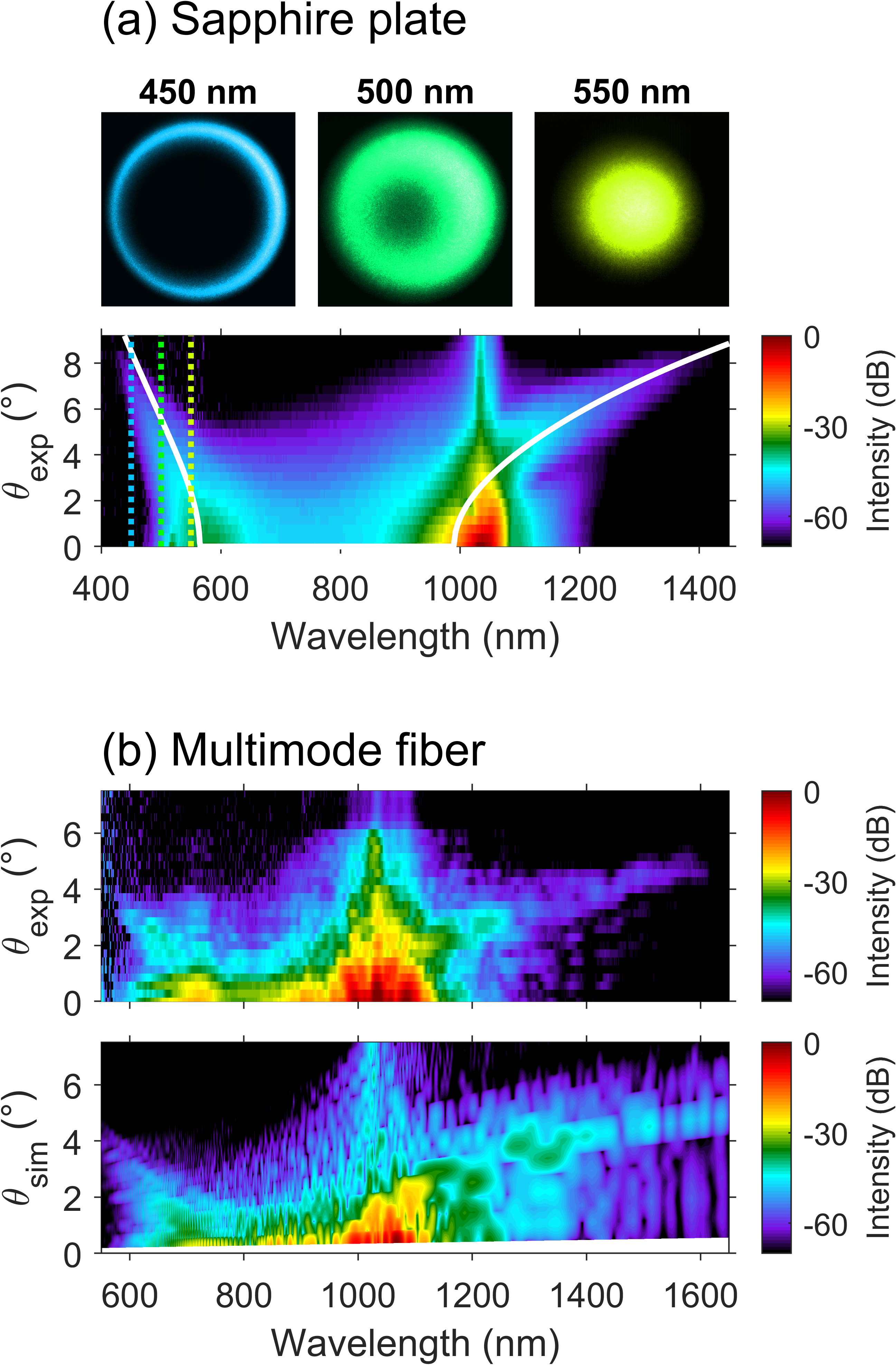}
    \caption{\label{fig:2}(a) Conical emission recorded at the output of the 5-mm-thick sapphire plate. Bottom panel: Measured angle-resolved far-field spectrum. Solid white line provides the corresponding phase-matching using relation of $\theta(\omega)$. Top panels: Near-field images recorded with the use of band-pass spectral filters whose central wavelengths are indicated with colored dashed lines in the bottom panel. (b) Top panel: Experimentally measured angle-resolved far-field spectrum at the output of our MMF segment. Bottom panel: Corresponding numerical simulation of angle-resolved far-field spectrum based on MM-UPPE model. Calculated angles take into account the refraction taking place at the fiber output. Note also that our spatial numerical grid cannot provide details for very small values of $\theta$.}
\end{figure}

In the following, we compare experimental results of conical emission in our MMF with our theoretical description and numerical modeling. The pulse energy coupled into the fiber was \SI{1.2}{\micro\joule}. First, a detailed analysis of conical emission patterns, shown in Fig.~\ref{fig:1}, is reported through measured angularly resolved spectra in Fig.~\ref{fig:2}. As an example of comparison, we measured the angle-resolved far-field spectrum ($\theta - \lambda$) obtained for the sapphire plate by using similar pulse energy and distinct focusing lens (see Fig.~\ref{fig:2}(a)). It exhibits the typical X-like structure studied in the normal dispersion regime of bulks \cite{faccio2007conical}. Beyond the continuous spectral broadening on-axis, the energy spreading follows an increasing cone angle of emission with increasing detuning from the carrier wavelength. This is confirmed by the different near-field images recorded with distinct band-pass spectral filters in the visible edge of the spectrum (see top panels in Fig.~\ref{fig:2}(a)). Regarding the interpretation of the full ($\theta - \lambda$) pattern observed, one can determine the central region of the angularly resolved spectrum where the energy of the dispersive waves gathers according to the phase-matching constraints (similarly to Eq.\,\ref{eq:conicalemission}). For bulk, the resulting relation simply writes as follows: $\theta(\omega)=\sin^{-1}\{c[k(\omega)^2-(K_0 + K_1 \omega)^2]^{1/2}/\omega\}$, where parameters $K_0$ and $K_1$ are related to phase and group velocities of the conical wave, and $k(\omega)=n(\omega)\omega/c$ with $n(\omega)$ the refractive index of the medium. Using a simple fitting procedure, one can retrieve the corresponding white lines shown in Fig.~\ref{fig:2}(a) and defining the theoretical phase-matching of X-wave. This situation is quite commonly observed where a spectral gap appears between the two X-tails, while one of these always passes through or close to $\omega_{0}$. This simply arises when the group velocity of the conical wave $1/K_1$ strictly differs from the one of the input pump pulse. 

In the top panel of Fig.~\ref{fig:2}(b), we unveil the angularly resolved spectrum recorded at the output of our short multimode fiber segment. We globally retrieve the typical scenario with strong spectral broadening on-axis and the emergence of hyperbolic X-like tails, clearly visible on both wavelength edges. Again, the spectral gap between the two X-tails can be easily noticed, thus confirming that the conical wave is not directly seeded by the input pump pulse. More details will be given in the following. Corresponding numerical simulation based on our MM-UPPE model confirms such a universal behaviour. As shown in the bottom panel of Fig.~\ref{fig:2}(b), an excellent agreement with experiment is obtained both in terms of angular emission and spectral bandwidth over a very high dynamic range. When comparing bulk and fiber patterns, we also note that maximum angle of emission and spectral edges of the spectral broadening differ. Phase-matching relations clearly reveal that such features are driven by the dispersion properties of the medium or waveguide as well as the group velocity of the seeding broadband localized wave structure. 
Moreover, it is worth mentioning that the ($\theta - \lambda$) or even the ($k_\perp - \omega$) representation is not the most suitable basis to describe the conical emission phenomenon occurring in a~waveguide. Even if we are not able to experimentally decompose the full output beam on the modal basis of our fiber, next we investigate the mode-resolved spectrum of conical emission.

\begin{figure*}
    \centering
    \includegraphics[width=0.7\linewidth]{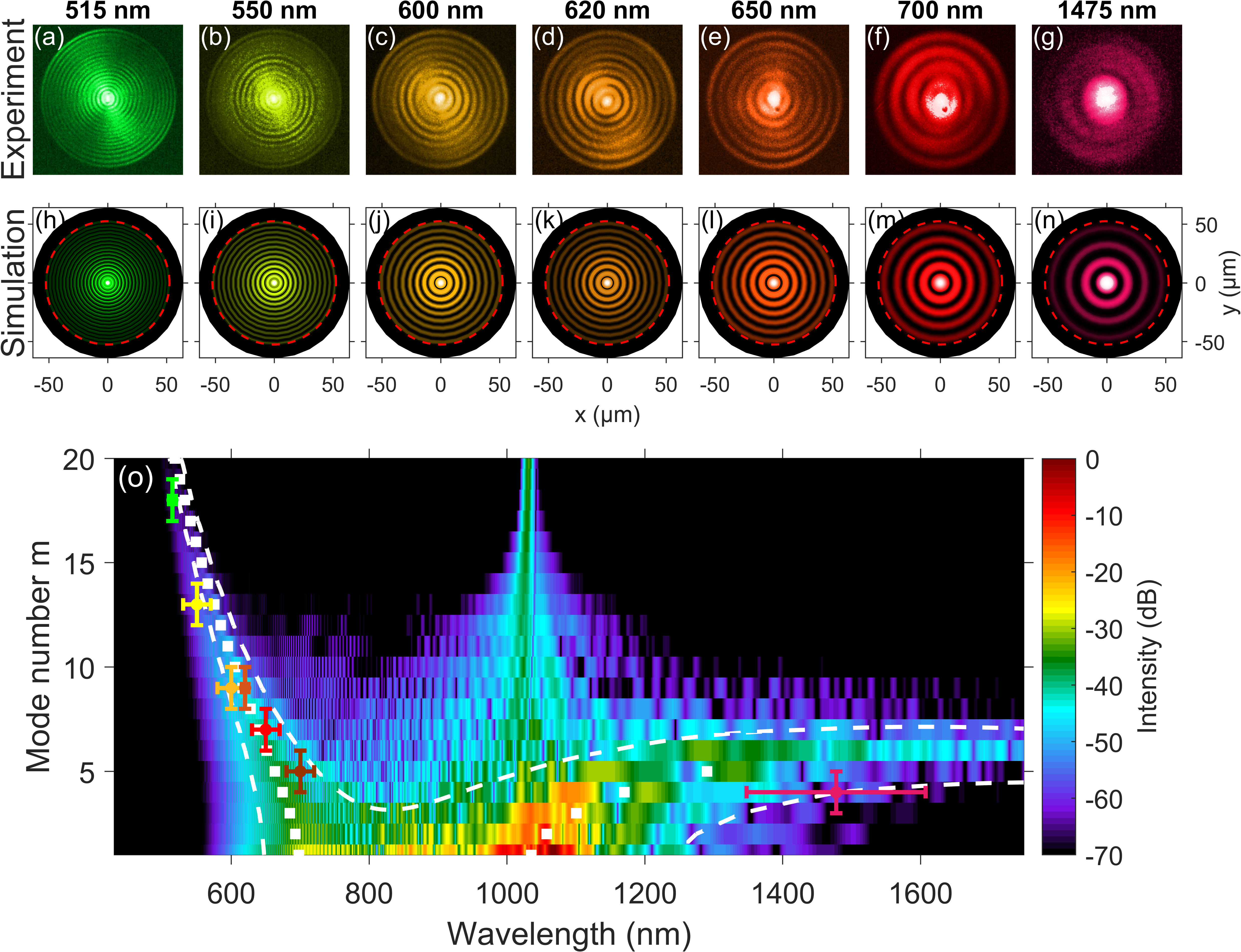}
    \caption{\label{fig:3}Mode-resolved spectrum of conical emission obtained at the output of our MMF segment. (a-n) Near-field images of the MMF higher-order modes involved in both X-like tails at specific wavelengths. Top panels: experimental captures with CCD camera. Bottom panels: corresponding numerical fiber modes. (o) False-color map of the numerical mode-resolved spectrum. White squares and dashed lines: theoretical phase-matching region of the discretized conical wave. Colored crosses indicate spectral positions of the higher-order modes captured in subplots (a-g) according to the distinct bandpass spectral filters.}
\end{figure*}

Figure~\ref{fig:3} shows the numerical results of the distribution of the full optical spectrum (power in log scale) over the different fiber modes after propagation. We still observe an evident X-shaped pattern of the field distribution that is very similar to those studied in the usual Fourier domain, except that the features of hyperbolic tails change on both wavelength edges (see Fig.~\ref{fig:3}(o)). The visible branch clearly exhibits higher-order modes than the infrared one, while in $\theta$ plane it was associated with smaller cone angles. Experimentally, we characterized the modal composition of both visible and infrared X-like tails by means of near-field imaging of the MMF output face combined with bandpass spectral filtering. Top panel in Fig.~\ref{fig:3} (subplots (a)-(g)) confirms the discrete nature of the conical emission. Increasingly-detuned frequencies from the pump are contained in higher-order modes of the MMF. In particular, for the visible tail, the phase-matching is narrow enough in $(m,\omega)$ plane so that each filtered spectral band (10-or 40-nm bandwidth) is nearly associated with only one or a few fiber modes. For instance, at \SI{515}{\nano\meter}, we are able to reveal the emission in LP$_{0,18}$ with a 10-\si{\nano\meter} bandwidth filter. And at \SI{550}{\nano\meter}, we can observe emission in LP$_{0,13}$ with a 40-\si{\nano\meter} bandwidth filter. By contrast, the infrared part of conical emission almost implies the same few higher-order modes (i.e., the tail is flatter beyond \SI{1200}{\nano\meter}), so that the modal analysis is less relevant. The main corresponding LP$_{0,m}$ modes, calculated with our scalar mode solver, are reported in Figs. 3(h-n) to corroborate the main modal content of our characterized X-like tails.

\begin{figure*}
    \centering
    \includegraphics[width=0.7\linewidth]{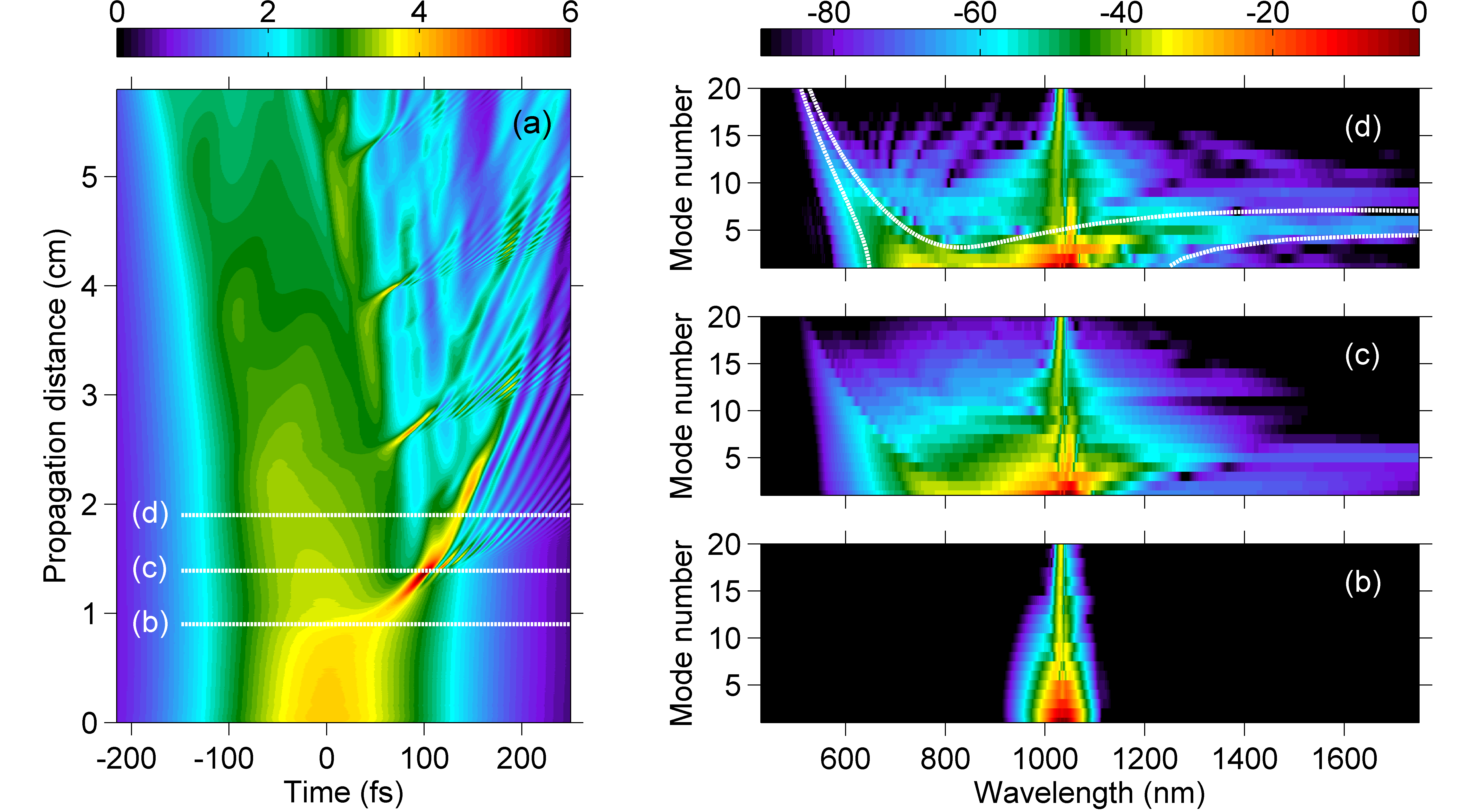}
    \caption{\label{fig:4}Evolution of the nonlinear pulse propagation. (a) False-color map of temporal power profile as a function of propagation distance $z$ (linear scale, MW units). (b-d) False-color map of the numerical mode-resolved spectrum (in log. scale, dB units) calculated at different distances $z = 0.9$, $1.39$, and $\SI{1.91}{\centi\meter}$, respectively. Corresponding locations are indicated by white dashed lines in subplot (a), namely, before, during, and after the formation of the most intense optical shock. White dashed lines in subplot (d) show theoretical phase-matching region of the discretized conical wave.}
\end{figure*}

Next, we unveil the main nonlinear dynamics in the ST domain at the origin of the conical emission. The detailed nonlinear pulse propagation in the time domain is shown in Fig.~\ref{fig:4}(a) whereas the corresponding evolution of mode-resolved spectrum is depicted in Fig.~\ref{fig:4}(b-d). In particular, three snapshots at distinct distances are provided, namely just before, during, and just after the conical emission. We observed that
the spectral dynamics almost reaches a stationary state after 2 cm, just past the first pulse splitting phenomenon
observed in the time domain (see Fig.~\ref{fig:4}(a)). This typical scenario already analyzed in normal dispersion of
bulk media occurs near the nonlinear spatiotemporal focus due to the self-steepening effect \cite{kibler2021discretized}. In particular, the velocity difference between the trailing pulse peak and its tails leads to the formation of an optical shock (at the trailing edge) due to the nonlinear dependence of the refractive index. The strong spectral broadening associated with shock formation at the trailing edge of the pulse clearly seeds a~large number of modes (see Fig.~\ref{fig:4}(c)), but only phase-matched dispersive radiations over specific higher-order modes remain with significant intensities (see Fig.~\ref{fig:4}(d)) according to the velocity of the shock front, as described by Eq.\,\ref{eq:conicalemission}.
More specifically, it was found that the difference in group velocities of the shock front and $v_\mathrm{g_{0}}$ is \SI{12.6}{\pico\second\per\meter} in Fig.~\ref{fig:4}(a), whereas the finite distance $d_z = \SI{1.4}{\milli\meter}$ (over which the seeding process occurs -- between $z = \SI{1.28}{\centi\meter}$ and $z = \SI{1.42}{\centi\meter}$) well defines the finite bandwidth of our theoretical phase-matching shown in Fig.~4(d), see white dashed lines. The Eq.\,\ref{eq:conicalemission} provides a generalized phase-matching rule of resonant radiations emitted by a broadband localized structure (soliton, shock front) through intra- and intermodal properties of any waveguide and whatever the dispersion regime. We can mention that the simplest and limiting case of an intermodal dispersive radiation emitted by a soliton pulse has been already studied in a few-mode fiber \cite{cheng2012intermodal}. A more generalized approach of dispersive wave emission has been recently derived for a multimode soliton \cite{eftekhar2021general}, similarly to our phase-matching relation. 
Here, our linear coherent superposition of radiations with an engineered spatiotemporal spectrum confers particular 2D+1 ST features to this wavepacket. Nevertheless, the spectral tolerance of the phase-matching in this spontaneous emergence of conical waves will lead to the dispersion of the conical wave \cite{kibler2021discretized}.

Note that several less-intense optical shocks on the trailing edge of the pump pulse occur after 2 cm in our simulation (see Fig.~\ref{fig:4}(a) at $z = \SI{2.7}{\centi\meter}$, $z = \SI{4.0}{\centi\meter}$ and $z = \SI{5.2}{\centi\meter}$). The quasi-periodicity of their occurrence with propagation distance can be easily linked to beating of the first two modes of the fiber (LP$_{0,1}$ and LP$_{0,2}$). Such processes then seed less-intense discretized conical waves with similar features in the mode-resolved spectrum, thus inducing some additional broadening of both X-like tails observed in the final map shown in Fig.~\ref{fig:3}(o). We also transposed the theoretical phase-matching obtained in Fig.~\ref{fig:4}(d) to Fig.~\ref{fig:3}(o), thus corroborating our experimental characterization of spontaneous conical emission.

\section{Conclusion}

In summary, we have experimentally demonstrated the spontaneous emission of a discretized conical wave when an intense ultrashort pulse propagates nonlinearly in a step-index multimode fiber, similarly to bulk conical emission. The resulting 2D+1 ST wavepacket propagates with a deterministic group-velocity and it results from a linear superposition of fiber modes with an engineered spatiotemporal spectrum. Even if the configuration of spontaneous emergence implies that the conical wave is not fully non-dispersive and non-diffractive, this simple and alternative approach opens the way to future extensive studies of multidimensional ST wavepackets with fiber and waveguide technologies \cite{shiri2022propagation,kolesik2009supercontinuum,walker2019spatiotemporal,wright2022}. Finally, our results also prove the general capabilities of MM-UPPE modeling for nonlinear multimode propagation problems.

\begin{acknowledgments}
The authors acknowledge financial support of French programs “Investments for the Future” operated by the National Research Agency (ISITE-BFC, contract ANR-15-IDEX-03; EIPHI Graduate School, contract ANR-17-EURE-0002; EQUIPEX+ Smartlight, contract ANR-21-ESRE-0040), from Bourgogne Franche-Comté region and European Regional Development Fund, from Ministère de l’Europe et des Affaires Étrangères, Ministère de l’Enseignement Superieur, de la Recherche et de l’Innovation (through French-Polish Polonium Hubert Curien Partnership), and from Narodowa Agencja Wymiany Akademickiej (Polonium 2019-2020) and Narodowe Centrum Nauki (2018/30/E/ST7/00862).
\end{acknowledgments}

\bibliography{apssamp}% Produces the bibliography via BibTeX.

\begin{thebibliography}{27}%
\makeatletter
\providecommand \@ifxundefined [1]{%
 \@ifx{#1\undefined}
}%
\providecommand \@ifnum [1]{%
 \ifnum #1\expandafter \@firstoftwo
 \else \expandafter \@secondoftwo
 \fi
}%
\providecommand \@ifx [1]{%
 \ifx #1\expandafter \@firstoftwo
 \else \expandafter \@secondoftwo
 \fi
}%
\providecommand \natexlab [1]{#1}%
\providecommand \enquote  [1]{``#1''}%
\providecommand \bibnamefont  [1]{#1}%
\providecommand \bibfnamefont [1]{#1}%
\providecommand \citenamefont [1]{#1}%
\providecommand \href@noop [0]{\@secondoftwo}%
\providecommand \href [0]{\begingroup \@sanitize@url \@href}%
\providecommand \@href[1]{\@@startlink{#1}\@@href}%
\providecommand \@@href[1]{\endgroup#1\@@endlink}%
\providecommand \@sanitize@url [0]{\catcode `\\12\catcode `\$12\catcode
  `\&12\catcode `\#12\catcode `\^12\catcode `\_12\catcode `\%12\relax}%
\providecommand \@@startlink[1]{}%
\providecommand \@@endlink[0]{}%
\providecommand \url  [0]{\begingroup\@sanitize@url \@url }%
\providecommand \@url [1]{\endgroup\@href {#1}{\urlprefix }}%
\providecommand \urlprefix  [0]{URL }%
\providecommand \Eprint [0]{\href }%
\providecommand \doibase [0]{https://doi.org/}%
\providecommand \selectlanguage [0]{\@gobble}%
\providecommand \bibinfo  [0]{\@secondoftwo}%
\providecommand \bibfield  [0]{\@secondoftwo}%
\providecommand \translation [1]{[#1]}%
\providecommand \BibitemOpen [0]{}%
\providecommand \bibitemStop [0]{}%
\providecommand \bibitemNoStop [0]{.\EOS\space}%
\providecommand \EOS [0]{\spacefactor3000\relax}%
\providecommand \BibitemShut  [1]{\csname bibitem#1\endcsname}%
\let\auto@bib@innerbib\@empty
%</preamble>
\bibitem [{\citenamefont {Picozzi}\ \emph {et~al.}(2015)\citenamefont
  {Picozzi}, \citenamefont {Millot},\ and\ \citenamefont
  {Wabnitz}}]{picozzi2015nonlinear}%
  \BibitemOpen
  \bibfield  {author} {\bibinfo {author} {\bibfnamefont {A.}~\bibnamefont
  {Picozzi}}, \bibinfo {author} {\bibfnamefont {G.}~\bibnamefont {Millot}},\
  and\ \bibinfo {author} {\bibfnamefont {S.}~\bibnamefont {Wabnitz}},\
  }\bibfield  {title} {\bibinfo {title} {Nonlinear virtues of multimode
  fibre},\ }\href@noop {} {\bibfield  {journal} {\bibinfo  {journal} {Nature
  Photonics}\ }\textbf {\bibinfo {volume} {9}},\ \bibinfo {pages} {289}
  (\bibinfo {year} {2015})}\BibitemShut {NoStop}%
\bibitem [{\citenamefont {Wright}\ \emph {et~al.}(2015)\citenamefont {Wright},
  \citenamefont {Christodoulides},\ and\ \citenamefont
  {Wise}}]{wright2015controllable}%
  \BibitemOpen
  \bibfield  {author} {\bibinfo {author} {\bibfnamefont {L.~G.}\ \bibnamefont
  {Wright}}, \bibinfo {author} {\bibfnamefont {D.~N.}\ \bibnamefont
  {Christodoulides}},\ and\ \bibinfo {author} {\bibfnamefont {F.~W.}\
  \bibnamefont {Wise}},\ }\bibfield  {title} {\bibinfo {title} {Controllable
  spatiotemporal nonlinear effects in multimode fibres},\ }\href@noop {}
  {\bibfield  {journal} {\bibinfo  {journal} {Nature photonics}\ }\textbf
  {\bibinfo {volume} {9}},\ \bibinfo {pages} {306} (\bibinfo {year}
  {2015})}\BibitemShut {NoStop}%
\bibitem [{\citenamefont {Krupa}\ \emph {et~al.}(2016)\citenamefont {Krupa},
  \citenamefont {Tonello}, \citenamefont {Barth{\'e}l{\'e}my}, \citenamefont
  {Couderc}, \citenamefont {Shalaby}, \citenamefont {Bendahmane}, \citenamefont
  {Millot},\ and\ \citenamefont {Wabnitz}}]{krupa2016observation}%
  \BibitemOpen
  \bibfield  {author} {\bibinfo {author} {\bibfnamefont {K.}~\bibnamefont
  {Krupa}}, \bibinfo {author} {\bibfnamefont {A.}~\bibnamefont {Tonello}},
  \bibinfo {author} {\bibfnamefont {A.}~\bibnamefont {Barth{\'e}l{\'e}my}},
  \bibinfo {author} {\bibfnamefont {V.}~\bibnamefont {Couderc}}, \bibinfo
  {author} {\bibfnamefont {B.~M.}\ \bibnamefont {Shalaby}}, \bibinfo {author}
  {\bibfnamefont {A.}~\bibnamefont {Bendahmane}}, \bibinfo {author}
  {\bibfnamefont {G.}~\bibnamefont {Millot}},\ and\ \bibinfo {author}
  {\bibfnamefont {S.}~\bibnamefont {Wabnitz}},\ }\bibfield  {title} {\bibinfo
  {title} {Observation of geometric parametric instability induced by the
  periodic spatial self-imaging of multimode waves},\ }\href@noop {} {\bibfield
   {journal} {\bibinfo  {journal} {Physical review letters}\ }\textbf {\bibinfo
  {volume} {116}},\ \bibinfo {pages} {183901} (\bibinfo {year}
  {2016})}\BibitemShut {NoStop}%
\bibitem [{\citenamefont {Krupa}\ \emph {et~al.}(2017)\citenamefont {Krupa},
  \citenamefont {Tonello}, \citenamefont {Shalaby}, \citenamefont {Fabert},
  \citenamefont {Barth{\'e}l{\'e}my}, \citenamefont {Millot}, \citenamefont
  {Wabnitz},\ and\ \citenamefont {Couderc}}]{krupa2017spatial}%
  \BibitemOpen
  \bibfield  {author} {\bibinfo {author} {\bibfnamefont {K.}~\bibnamefont
  {Krupa}}, \bibinfo {author} {\bibfnamefont {A.}~\bibnamefont {Tonello}},
  \bibinfo {author} {\bibfnamefont {B.~M.}\ \bibnamefont {Shalaby}}, \bibinfo
  {author} {\bibfnamefont {M.}~\bibnamefont {Fabert}}, \bibinfo {author}
  {\bibfnamefont {A.}~\bibnamefont {Barth{\'e}l{\'e}my}}, \bibinfo {author}
  {\bibfnamefont {G.}~\bibnamefont {Millot}}, \bibinfo {author} {\bibfnamefont
  {S.}~\bibnamefont {Wabnitz}},\ and\ \bibinfo {author} {\bibfnamefont
  {V.}~\bibnamefont {Couderc}},\ }\bibfield  {title} {\bibinfo {title} {Spatial
  beam self-cleaning in multimode fibres},\ }\href@noop {} {\bibfield
  {journal} {\bibinfo  {journal} {Nature Photonics}\ }\textbf {\bibinfo
  {volume} {11}},\ \bibinfo {pages} {237} (\bibinfo {year} {2017})}\BibitemShut
  {NoStop}%
\bibitem [{\citenamefont {Renninger}\ and\ \citenamefont
  {Wise}(2013)}]{renninger2013optical}%
  \BibitemOpen
  \bibfield  {author} {\bibinfo {author} {\bibfnamefont {W.~H.}\ \bibnamefont
  {Renninger}}\ and\ \bibinfo {author} {\bibfnamefont {F.~W.}\ \bibnamefont
  {Wise}},\ }\bibfield  {title} {\bibinfo {title} {Optical solitons in
  graded-index multimode fibres},\ }\href@noop {} {\bibfield  {journal}
  {\bibinfo  {journal} {Nature communications}\ }\textbf {\bibinfo {volume}
  {4}},\ \bibinfo {pages} {1} (\bibinfo {year} {2013})}\BibitemShut {NoStop}%
\bibitem [{\citenamefont {Wright}\ \emph {et~al.}(2017)\citenamefont {Wright},
  \citenamefont {Christodoulides},\ and\ \citenamefont
  {Wise}}]{wright2017spatiotemporal}%
  \BibitemOpen
  \bibfield  {author} {\bibinfo {author} {\bibfnamefont {L.~G.}\ \bibnamefont
  {Wright}}, \bibinfo {author} {\bibfnamefont {D.~N.}\ \bibnamefont
  {Christodoulides}},\ and\ \bibinfo {author} {\bibfnamefont {F.~W.}\
  \bibnamefont {Wise}},\ }\bibfield  {title} {\bibinfo {title} {Spatiotemporal
  mode-locking in multimode fiber lasers},\ }\href@noop {} {\bibfield
  {journal} {\bibinfo  {journal} {Science}\ }\textbf {\bibinfo {volume}
  {358}},\ \bibinfo {pages} {94} (\bibinfo {year} {2017})}\BibitemShut
  {NoStop}%
\bibitem [{\citenamefont {Pourbeyram}\ \emph {et~al.}(2022)\citenamefont
  {Pourbeyram}, \citenamefont {Sidorenko}, \citenamefont {Wu}, \citenamefont
  {Bender}, \citenamefont {Wright}, \citenamefont {Christodoulides},\ and\
  \citenamefont {Wise}}]{pourbeyram2022direct}%
  \BibitemOpen
  \bibfield  {author} {\bibinfo {author} {\bibfnamefont {H.}~\bibnamefont
  {Pourbeyram}}, \bibinfo {author} {\bibfnamefont {P.}~\bibnamefont
  {Sidorenko}}, \bibinfo {author} {\bibfnamefont {F.~O.}\ \bibnamefont {Wu}},
  \bibinfo {author} {\bibfnamefont {N.}~\bibnamefont {Bender}}, \bibinfo
  {author} {\bibfnamefont {L.}~\bibnamefont {Wright}}, \bibinfo {author}
  {\bibfnamefont {D.~N.}\ \bibnamefont {Christodoulides}},\ and\ \bibinfo
  {author} {\bibfnamefont {F.}~\bibnamefont {Wise}},\ }\bibfield  {title}
  {\bibinfo {title} {Direct observations of thermalization to a rayleigh--jeans
  distribution in multimode optical fibres},\ }\href@noop {} {\bibfield
  {journal} {\bibinfo  {journal} {Nature Physics}\ ,\ \bibinfo {pages} {1}}
  (\bibinfo {year} {2022})}\BibitemShut {NoStop}%
\bibitem [{\citenamefont {Podivilov}\ \emph {et~al.}(2022)\citenamefont
  {Podivilov}, \citenamefont {Mangini}, \citenamefont {Sidelnikov},
  \citenamefont {Ferraro}, \citenamefont {Gervaziev}, \citenamefont {Kharenko},
  \citenamefont {Zitelli}, \citenamefont {Fedoruk}, \citenamefont {Babin},\
  and\ \citenamefont {Wabnitz}}]{podivilov2022thermalization}%
  \BibitemOpen
  \bibfield  {author} {\bibinfo {author} {\bibfnamefont {E.}~\bibnamefont
  {Podivilov}}, \bibinfo {author} {\bibfnamefont {F.}~\bibnamefont {Mangini}},
  \bibinfo {author} {\bibfnamefont {O.}~\bibnamefont {Sidelnikov}}, \bibinfo
  {author} {\bibfnamefont {M.}~\bibnamefont {Ferraro}}, \bibinfo {author}
  {\bibfnamefont {M.}~\bibnamefont {Gervaziev}}, \bibinfo {author}
  {\bibfnamefont {D.}~\bibnamefont {Kharenko}}, \bibinfo {author}
  {\bibfnamefont {M.}~\bibnamefont {Zitelli}}, \bibinfo {author} {\bibfnamefont
  {M.}~\bibnamefont {Fedoruk}}, \bibinfo {author} {\bibfnamefont
  {S.}~\bibnamefont {Babin}},\ and\ \bibinfo {author} {\bibfnamefont
  {S.}~\bibnamefont {Wabnitz}},\ }\bibfield  {title} {\bibinfo {title}
  {Thermalization of orbital angular momentum beams in multimode optical
  fibers},\ }\href@noop {} {\bibfield  {journal} {\bibinfo  {journal} {Physical
  Review Letters}\ }\textbf {\bibinfo {volume} {128}},\ \bibinfo {pages}
  {243901} (\bibinfo {year} {2022})}\BibitemShut {NoStop}%
\bibitem [{\citenamefont {Kibler}\ and\ \citenamefont
  {B{\'e}jot}(2021)}]{kibler2021discretized}%
  \BibitemOpen
  \bibfield  {author} {\bibinfo {author} {\bibfnamefont {B.}~\bibnamefont
  {Kibler}}\ and\ \bibinfo {author} {\bibfnamefont {P.}~\bibnamefont
  {B{\'e}jot}},\ }\bibfield  {title} {\bibinfo {title} {Discretized conical
  waves in multimode optical fibers},\ }\href@noop {} {\bibfield  {journal}
  {\bibinfo  {journal} {Physical Review Letters}\ }\textbf {\bibinfo {volume}
  {126}},\ \bibinfo {pages} {023902} (\bibinfo {year} {2021})}\BibitemShut
  {NoStop}%
\bibitem [{\citenamefont {B{\'e}jot}\ and\ \citenamefont
  {Kibler}(2021)}]{bejot2021spatiotemporal}%
  \BibitemOpen
  \bibfield  {author} {\bibinfo {author} {\bibfnamefont {P.}~\bibnamefont
  {B{\'e}jot}}\ and\ \bibinfo {author} {\bibfnamefont {B.}~\bibnamefont
  {Kibler}},\ }\bibfield  {title} {\bibinfo {title} {Spatiotemporal helicon
  wavepackets},\ }\href@noop {} {\bibfield  {journal} {\bibinfo  {journal} {ACS
  photonics}\ }\textbf {\bibinfo {volume} {8}},\ \bibinfo {pages} {2345}
  (\bibinfo {year} {2021})}\BibitemShut {NoStop}%
\bibitem [{\citenamefont {Shiri}\ \emph {et~al.}(2020)\citenamefont {Shiri},
  \citenamefont {Yessenov}, \citenamefont {Webster}, \citenamefont {Schepler},\
  and\ \citenamefont {Abouraddy}}]{shiri2020hybrid}%
  \BibitemOpen
  \bibfield  {author} {\bibinfo {author} {\bibfnamefont {A.}~\bibnamefont
  {Shiri}}, \bibinfo {author} {\bibfnamefont {M.}~\bibnamefont {Yessenov}},
  \bibinfo {author} {\bibfnamefont {S.}~\bibnamefont {Webster}}, \bibinfo
  {author} {\bibfnamefont {K.~L.}\ \bibnamefont {Schepler}},\ and\ \bibinfo
  {author} {\bibfnamefont {A.~F.}\ \bibnamefont {Abouraddy}},\ }\bibfield
  {title} {\bibinfo {title} {Hybrid guided space-time optical modes in
  unpatterned films},\ }\href@noop {} {\bibfield  {journal} {\bibinfo
  {journal} {Nature Communications}\ }\textbf {\bibinfo {volume} {11}},\
  \bibinfo {pages} {1} (\bibinfo {year} {2020})}\BibitemShut {NoStop}%
\bibitem [{\citenamefont {B{\'e}jot}\ and\ \citenamefont
  {Kibler}(2022)}]{bejot2022quadrics}%
  \BibitemOpen
  \bibfield  {author} {\bibinfo {author} {\bibfnamefont {P.}~\bibnamefont
  {B{\'e}jot}}\ and\ \bibinfo {author} {\bibfnamefont {B.}~\bibnamefont
  {Kibler}},\ }\bibfield  {title} {\bibinfo {title} {Quadrics for structuring
  invariant space--time wavepackets},\ }\href@noop {} {\bibfield  {journal}
  {\bibinfo  {journal} {ACS Photonics}\ } (\bibinfo {year} {2022})}\BibitemShut
  {NoStop}%
\bibitem [{\citenamefont {Shiri}\ \emph {et~al.}(2022)\citenamefont {Shiri},
  \citenamefont {Webster}, \citenamefont {Schepler},\ and\ \citenamefont
  {Abouraddy}}]{shiri2022propagation}%
  \BibitemOpen
  \bibfield  {author} {\bibinfo {author} {\bibfnamefont {A.}~\bibnamefont
  {Shiri}}, \bibinfo {author} {\bibfnamefont {S.}~\bibnamefont {Webster}},
  \bibinfo {author} {\bibfnamefont {K.~L.}\ \bibnamefont {Schepler}},\ and\
  \bibinfo {author} {\bibfnamefont {A.~F.}\ \bibnamefont {Abouraddy}},\
  }\bibfield  {title} {\bibinfo {title} {Propagation-invariant space-time
  supermodes in a multimode waveguide},\ }\href@noop {} {\bibfield  {journal}
  {\bibinfo  {journal} {arXiv preprint arXiv:2204.01867}\ } (\bibinfo {year}
  {2022})}\BibitemShut {NoStop}%
\bibitem [{\citenamefont {Kondakci}\ and\ \citenamefont
  {Abouraddy}(2019)}]{kondakci2019optical}%
  \BibitemOpen
  \bibfield  {author} {\bibinfo {author} {\bibfnamefont {H.}~\bibnamefont
  {Kondakci}}\ and\ \bibinfo {author} {\bibfnamefont {A.~F.}\ \bibnamefont
  {Abouraddy}},\ }\bibfield  {title} {\bibinfo {title} {Optical space-time wave
  packets having arbitrary group velocities in free space},\ }\href@noop {}
  {\bibfield  {journal} {\bibinfo  {journal} {Nature communications}\ }\textbf
  {\bibinfo {volume} {10}},\ \bibinfo {pages} {1} (\bibinfo {year}
  {2019})}\BibitemShut {NoStop}%
\bibitem [{\citenamefont {Yessenov}\ \emph {et~al.}(2021)\citenamefont
  {Yessenov}, \citenamefont {Free}, \citenamefont {Chen}, \citenamefont
  {Johnson}, \citenamefont {Lavery}, \citenamefont {Alonso},\ and\
  \citenamefont {Abouraddy}}]{yessenov2021space}%
  \BibitemOpen
  \bibfield  {author} {\bibinfo {author} {\bibfnamefont {M.}~\bibnamefont
  {Yessenov}}, \bibinfo {author} {\bibfnamefont {J.}~\bibnamefont {Free}},
  \bibinfo {author} {\bibfnamefont {Z.}~\bibnamefont {Chen}}, \bibinfo {author}
  {\bibfnamefont {E.~G.}\ \bibnamefont {Johnson}}, \bibinfo {author}
  {\bibfnamefont {M.~P.}\ \bibnamefont {Lavery}}, \bibinfo {author}
  {\bibfnamefont {M.~A.}\ \bibnamefont {Alonso}},\ and\ \bibinfo {author}
  {\bibfnamefont {A.~F.}\ \bibnamefont {Abouraddy}},\ }\bibfield  {title}
  {\bibinfo {title} {Space-time wave packets localized in all dimensions},\
  }\href@noop {} {\bibfield  {journal} {\bibinfo  {journal} {arXiv preprint
  arXiv:2111.03095}\ } (\bibinfo {year} {2021})}\BibitemShut {NoStop}%
\bibitem [{\citenamefont {Yessenov}\ \emph {et~al.}(2022)\citenamefont
  {Yessenov}, \citenamefont {Hall}, \citenamefont {Schepler},\ and\
  \citenamefont {Abouraddy}}]{yessenov2022space}%
  \BibitemOpen
  \bibfield  {author} {\bibinfo {author} {\bibfnamefont {M.}~\bibnamefont
  {Yessenov}}, \bibinfo {author} {\bibfnamefont {L.~A.}\ \bibnamefont {Hall}},
  \bibinfo {author} {\bibfnamefont {K.~L.}\ \bibnamefont {Schepler}},\ and\
  \bibinfo {author} {\bibfnamefont {A.~F.}\ \bibnamefont {Abouraddy}},\
  }\bibfield  {title} {\bibinfo {title} {Space-time wave packets},\ }\href@noop
  {} {\bibfield  {journal} {\bibinfo  {journal} {arXiv preprint
  arXiv:2201.08297}\ } (\bibinfo {year} {2022})}\BibitemShut {NoStop}%
\bibitem [{\citenamefont {Faccio}\ \emph {et~al.}(2007)\citenamefont {Faccio},
  \citenamefont {Couairon},\ and\ \citenamefont
  {Di~Trapani}}]{faccio2007conical}%
  \BibitemOpen
  \bibfield  {author} {\bibinfo {author} {\bibfnamefont {D.}~\bibnamefont
  {Faccio}}, \bibinfo {author} {\bibfnamefont {A.}~\bibnamefont {Couairon}},\
  and\ \bibinfo {author} {\bibfnamefont {P.}~\bibnamefont {Di~Trapani}},\
  }\href@noop {} {\emph {\bibinfo {title} {Conical waves, filaments and
  nonlinear filamentation optics}}}\ (\bibinfo  {publisher} {Aracne Rome},\
  \bibinfo {year} {2007})\BibitemShut {NoStop}%
\bibitem [{\citenamefont {Hern{\'a}ndez-Figueroa}\ \emph
  {et~al.}(2008)\citenamefont {Hern{\'a}ndez-Figueroa}, \citenamefont
  {Zamboni-Rached},\ and\ \citenamefont {Recami}}]{hernandez2008localized}%
  \BibitemOpen
  \bibfield  {author} {\bibinfo {author} {\bibfnamefont {H.~E.}\ \bibnamefont
  {Hern{\'a}ndez-Figueroa}}, \bibinfo {author} {\bibfnamefont {M.}~\bibnamefont
  {Zamboni-Rached}},\ and\ \bibinfo {author} {\bibfnamefont {E.}~\bibnamefont
  {Recami}},\ }\href@noop {} {\emph {\bibinfo {title} {Localized waves}}},\
  Vol.\ \bibinfo {volume} {194}\ (\bibinfo  {publisher} {John Wiley \& Sons},\
  \bibinfo {year} {2008})\BibitemShut {NoStop}%
\bibitem [{\citenamefont {B{\'e}jot}(2019)}]{bejot2019multimodal}%
  \BibitemOpen
  \bibfield  {author} {\bibinfo {author} {\bibfnamefont {P.}~\bibnamefont
  {B{\'e}jot}},\ }\bibfield  {title} {\bibinfo {title} {Multimodal
  unidirectional pulse propagation equation},\ }\href@noop {} {\bibfield
  {journal} {\bibinfo  {journal} {Physical Review E}\ }\textbf {\bibinfo
  {volume} {99}},\ \bibinfo {pages} {032217} (\bibinfo {year}
  {2019})}\BibitemShut {NoStop}%
\bibitem [{\citenamefont {Tarnowski}\ \emph {et~al.}(2021)\citenamefont
  {Tarnowski}, \citenamefont {Majchrowska}, \citenamefont {B{\'e}jot},\ and\
  \citenamefont {Kibler}}]{tarnowski2021numerical}%
  \BibitemOpen
  \bibfield  {author} {\bibinfo {author} {\bibfnamefont {K.}~\bibnamefont
  {Tarnowski}}, \bibinfo {author} {\bibfnamefont {S.}~\bibnamefont
  {Majchrowska}}, \bibinfo {author} {\bibfnamefont {P.}~\bibnamefont
  {B{\'e}jot}},\ and\ \bibinfo {author} {\bibfnamefont {B.}~\bibnamefont
  {Kibler}},\ }\bibfield  {title} {\bibinfo {title} {Numerical modelings of
  ultrashort pulse propagation and conical emission in multimode optical
  fibers},\ }\href@noop {} {\bibfield  {journal} {\bibinfo  {journal} {JOSA B}\
  }\textbf {\bibinfo {volume} {38}},\ \bibinfo {pages} {732} (\bibinfo {year}
  {2021})}\BibitemShut {NoStop}%
\bibitem [{\citenamefont {Kolesik}\ \emph {et~al.}(2005)\citenamefont
  {Kolesik}, \citenamefont {Wright},\ and\ \citenamefont
  {Moloney}}]{kolesik2005interpretation}%
  \BibitemOpen
  \bibfield  {author} {\bibinfo {author} {\bibfnamefont {M.}~\bibnamefont
  {Kolesik}}, \bibinfo {author} {\bibfnamefont {E.~M.}\ \bibnamefont
  {Wright}},\ and\ \bibinfo {author} {\bibfnamefont {J.~V.}\ \bibnamefont
  {Moloney}},\ }\bibfield  {title} {\bibinfo {title} {Interpretation of the
  spectrally resolved far field of femtosecond pulses propagating in bulk
  nonlinear dispersive media},\ }\href@noop {} {\bibfield  {journal} {\bibinfo
  {journal} {Optics express}\ }\textbf {\bibinfo {volume} {13}},\ \bibinfo
  {pages} {10729} (\bibinfo {year} {2005})}\BibitemShut {NoStop}%
\bibitem [{\citenamefont {Ferraro}\ \emph {et~al.}(2022)\citenamefont
  {Ferraro}, \citenamefont {Mangini}, \citenamefont {Sun}, \citenamefont
  {Zitelli}, \citenamefont {Niang}, \citenamefont {Crocco}, \citenamefont
  {Formoso}, \citenamefont {Agostino}, \citenamefont {Barberi}, \citenamefont
  {De~Luca} \emph {et~al.}}]{ferraro2022multiphoton}%
  \BibitemOpen
  \bibfield  {author} {\bibinfo {author} {\bibfnamefont {M.}~\bibnamefont
  {Ferraro}}, \bibinfo {author} {\bibfnamefont {F.}~\bibnamefont {Mangini}},
  \bibinfo {author} {\bibfnamefont {Y.}~\bibnamefont {Sun}}, \bibinfo {author}
  {\bibfnamefont {M.}~\bibnamefont {Zitelli}}, \bibinfo {author} {\bibfnamefont
  {A.}~\bibnamefont {Niang}}, \bibinfo {author} {\bibfnamefont {M.~C.}\
  \bibnamefont {Crocco}}, \bibinfo {author} {\bibfnamefont {V.}~\bibnamefont
  {Formoso}}, \bibinfo {author} {\bibfnamefont {R.}~\bibnamefont {Agostino}},
  \bibinfo {author} {\bibfnamefont {R.}~\bibnamefont {Barberi}}, \bibinfo
  {author} {\bibfnamefont {A.}~\bibnamefont {De~Luca}}, \emph {et~al.},\
  }\bibfield  {title} {\bibinfo {title} {Multiphoton ionization of standard
  optical fibers},\ }\href@noop {} {\bibfield  {journal} {\bibinfo  {journal}
  {Photonics Research}\ }\textbf {\bibinfo {volume} {10}},\ \bibinfo {pages}
  {1394} (\bibinfo {year} {2022})}\BibitemShut {NoStop}%
\bibitem [{\citenamefont {Cheng}\ \emph {et~al.}(2012)\citenamefont {Cheng},
  \citenamefont {Pedersen}, \citenamefont {Charan}, \citenamefont {Wang},
  \citenamefont {Xu}, \citenamefont {Gr{\"u}ner-Nielsen},\ and\ \citenamefont
  {Jakobsen}}]{cheng2012intermodal}%
  \BibitemOpen
  \bibfield  {author} {\bibinfo {author} {\bibfnamefont {J.}~\bibnamefont
  {Cheng}}, \bibinfo {author} {\bibfnamefont {M.~E.}\ \bibnamefont {Pedersen}},
  \bibinfo {author} {\bibfnamefont {K.}~\bibnamefont {Charan}}, \bibinfo
  {author} {\bibfnamefont {K.}~\bibnamefont {Wang}}, \bibinfo {author}
  {\bibfnamefont {C.}~\bibnamefont {Xu}}, \bibinfo {author} {\bibfnamefont
  {L.}~\bibnamefont {Gr{\"u}ner-Nielsen}},\ and\ \bibinfo {author}
  {\bibfnamefont {D.}~\bibnamefont {Jakobsen}},\ }\bibfield  {title} {\bibinfo
  {title} {Intermodal {\v{c}}erenkov radiation in a higher-order-mode fiber},\
  }\href@noop {} {\bibfield  {journal} {\bibinfo  {journal} {Optics letters}\
  }\textbf {\bibinfo {volume} {37}},\ \bibinfo {pages} {4410} (\bibinfo {year}
  {2012})}\BibitemShut {NoStop}%
\bibitem [{\citenamefont {Eftekhar}\ \emph {et~al.}(2021)\citenamefont
  {Eftekhar}, \citenamefont {Lopez-Aviles}, \citenamefont {Wise}, \citenamefont
  {Amezcua-Correa},\ and\ \citenamefont
  {Christodoulides}}]{eftekhar2021general}%
  \BibitemOpen
  \bibfield  {author} {\bibinfo {author} {\bibfnamefont {M.}~\bibnamefont
  {Eftekhar}}, \bibinfo {author} {\bibfnamefont {H.}~\bibnamefont
  {Lopez-Aviles}}, \bibinfo {author} {\bibfnamefont {F.}~\bibnamefont {Wise}},
  \bibinfo {author} {\bibfnamefont {R.}~\bibnamefont {Amezcua-Correa}},\ and\
  \bibinfo {author} {\bibfnamefont {D.}~\bibnamefont {Christodoulides}},\
  }\bibfield  {title} {\bibinfo {title} {General theory and observation of
  cherenkov radiation induced by multimode solitons},\ }\href@noop {}
  {\bibfield  {journal} {\bibinfo  {journal} {Communications Physics}\ }\textbf
  {\bibinfo {volume} {4}},\ \bibinfo {pages} {1} (\bibinfo {year}
  {2021})}\BibitemShut {NoStop}%
\bibitem [{\citenamefont {Kolesik}\ \emph {et~al.}(2009)\citenamefont
  {Kolesik}, \citenamefont {Faccio}, \citenamefont {Wright}, \citenamefont
  {Di~Trapani},\ and\ \citenamefont {Moloney}}]{kolesik2009supercontinuum}%
  \BibitemOpen
  \bibfield  {author} {\bibinfo {author} {\bibfnamefont {M.}~\bibnamefont
  {Kolesik}}, \bibinfo {author} {\bibfnamefont {D.}~\bibnamefont {Faccio}},
  \bibinfo {author} {\bibfnamefont {E.}~\bibnamefont {Wright}}, \bibinfo
  {author} {\bibfnamefont {P.}~\bibnamefont {Di~Trapani}},\ and\ \bibinfo
  {author} {\bibfnamefont {J.}~\bibnamefont {Moloney}},\ }\bibfield  {title}
  {\bibinfo {title} {Supercontinuum generation in planar glass membrane fibers:
  comparison with bulk media},\ }\href@noop {} {\bibfield  {journal} {\bibinfo
  {journal} {Optics letters}\ }\textbf {\bibinfo {volume} {34}},\ \bibinfo
  {pages} {286} (\bibinfo {year} {2009})}\BibitemShut {NoStop}%
\bibitem [{\citenamefont {Walker}\ \emph {et~al.}(2019)\citenamefont {Walker},
  \citenamefont {Whittaker}, \citenamefont {Skryabin}, \citenamefont
  {Cancellieri}, \citenamefont {Royall}, \citenamefont {Sich}, \citenamefont
  {Farrer}, \citenamefont {Ritchie}, \citenamefont {Skolnick},\ and\
  \citenamefont {Krizhanovskii}}]{walker2019spatiotemporal}%
  \BibitemOpen
  \bibfield  {author} {\bibinfo {author} {\bibfnamefont {P.~M.}\ \bibnamefont
  {Walker}}, \bibinfo {author} {\bibfnamefont {C.~E.}\ \bibnamefont
  {Whittaker}}, \bibinfo {author} {\bibfnamefont {D.~V.}\ \bibnamefont
  {Skryabin}}, \bibinfo {author} {\bibfnamefont {E.}~\bibnamefont
  {Cancellieri}}, \bibinfo {author} {\bibfnamefont {B.}~\bibnamefont {Royall}},
  \bibinfo {author} {\bibfnamefont {M.}~\bibnamefont {Sich}}, \bibinfo {author}
  {\bibfnamefont {I.}~\bibnamefont {Farrer}}, \bibinfo {author} {\bibfnamefont
  {D.~A.}\ \bibnamefont {Ritchie}}, \bibinfo {author} {\bibfnamefont {M.~S.}\
  \bibnamefont {Skolnick}},\ and\ \bibinfo {author} {\bibfnamefont {D.~N.}\
  \bibnamefont {Krizhanovskii}},\ }\bibfield  {title} {\bibinfo {title}
  {Spatiotemporal continuum generation in polariton waveguides},\ }\href@noop
  {} {\bibfield  {journal} {\bibinfo  {journal} {Light: Science \&
  Applications}\ }\textbf {\bibinfo {volume} {8}},\ \bibinfo {pages} {1}
  (\bibinfo {year} {2019})}\BibitemShut {NoStop}%
\bibitem [{\citenamefont {Wright}\ \emph {et~al.}(2022)\citenamefont {Wright},
  \citenamefont {Renninger}, \citenamefont {Christodoulides},\ and\
  \citenamefont {Wise}}]{wright2022}%
  \BibitemOpen
  \bibfield  {author} {\bibinfo {author} {\bibfnamefont {L.}~\bibnamefont
  {Wright}}, \bibinfo {author} {\bibfnamefont {W.}~\bibnamefont {Renninger}},
  \bibinfo {author} {\bibfnamefont {D.}~\bibnamefont {Christodoulides}},\ and\
  \bibinfo {author} {\bibfnamefont {F.}~\bibnamefont {Wise}},\ }\bibfield
  {title} {\bibinfo {title} {Nonlinear multimode photonics: nonlinear optics
  with many degrees of freedom},\ }\href@noop {} {\bibfield  {journal}
  {\bibinfo  {journal} {Optica}\ }\textbf {\bibinfo {volume} {9}},\ \bibinfo
  {pages} {824} (\bibinfo {year} {2022})}\BibitemShut {NoStop}%
\end{thebibliography}%

\end{document}